# Van der Waals heterostructures

A. K. Geim[1,2] & I. V. Grigorieva[1]

School of Physics & Astronomy[1] & Centre for Mesoscience & Nanotechnology[2]
University of Manchester, Manchester M13 9PL, United Kingdom

**Research on graphene and other two-dimensional atomic crystals is intense and likely to remain one of the hottest topics in condensed matter physics and materials science for many years. Looking beyond this field, isolated atomic planes can also be reassembled into designer heterostructures made layer by layer in a precisely chosen sequence. The first – already remarkably complex – such heterostructures (referred to as 'van der Waals') have recently been fabricated and investigated revealing unusual properties and new phenomena. Here we review this emerging research area and attempt to identify future directions. With steady improvement in fabrication techniques, van der Waals heterostructures promise a new gold rush, rather than a graphene aftershock.**

Graphene research has evolved into a vast field with approximately 10,000 papers now being published every year on a wide range of topics. Each topic is covered by many reviews. It is probably fair to say that research on 'simple graphene' has already passed its zenith. Indeed, the focus has shifted from studying graphene itself to the use of the material in applications[1] and as a versatile platform for investigation of various phenomena. Nonetheless, the fundamental science of graphene remains far from being exhausted (especially, in terms of many-body physics) and, as the quality of graphene devices continues to improve[2-5], more breakthroughs are expected although at a slower pace.

Because most 'low-hanging graphene fruits' have already been harvested, researchers have now started paying more attention to other two-dimensional (2D) atomic crystals[6] such as isolated mono- and few- layers of hexagonal boron nitride (hBN), molybdenum disulphide, other dichalcogenides and layered oxides. During the first 5 years of the graphene boom, there appeared only a few experimental papers on 2D crystals other than graphene, whereas the last two years have already seen many reviews (e.g., refs. 7-11). This research promises to reach the same intensity as that on graphene, especially if the electronic quality of 2D crystals such as $MoS_2$[12,13] can be improved by a factor of 10–100.

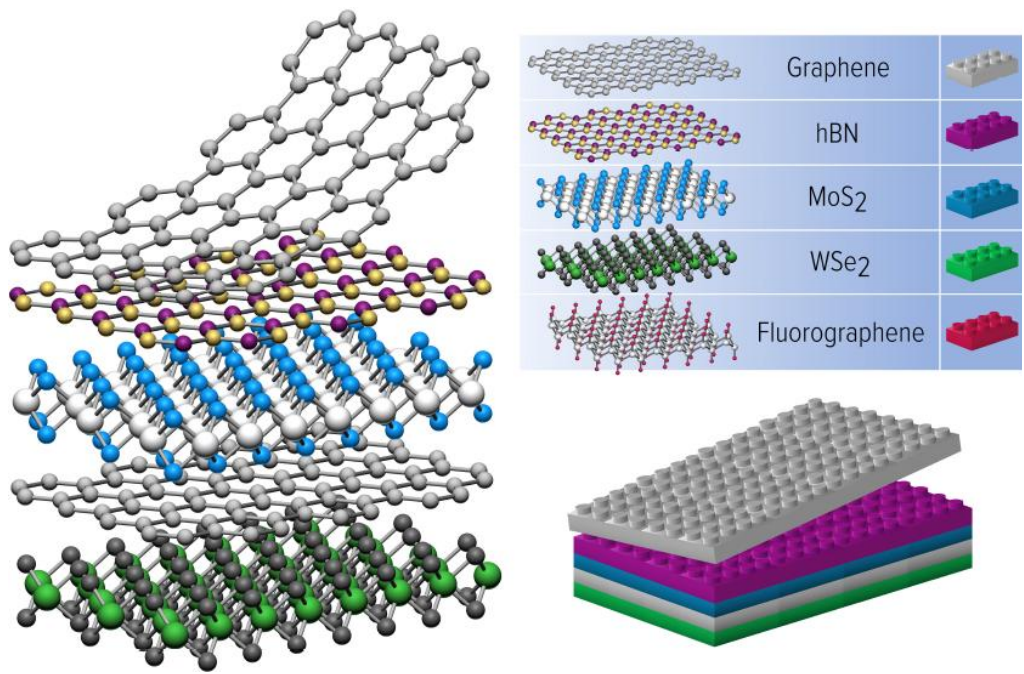

Fig. 1. **Building vdW heterostructures.** If one considers 2D crystals as Lego blocks (right panel), construction of a huge variety of layered structures becomes possible.  Conceptually, this atomic-scale Lego resembles molecular beam epitaxy but employs different 'construction' rules and a distinct set of materials.

In parallel with the efforts on graphene-like materials, another research field has recently emerged and has been gaining strength over the last two years. It deals with heterostructures and devices made by stacking different 2D crystals on top of each other. The basic principle is simple: take, for example, a monolayer, put it on top of another mono- or few- layer crystal, add another 2D crystal and so on. The resulting stack represents an artificial material assembled in a chosen sequence – like in a LEGO game – with blocks defined with one atomic plane precision (Fig. 1). Strong covalent bonds provide in-plane stability of 2D crystals whereas relatively weak, van der Waals-like forces are sufficient to keep the stack together. The possibility of making multilayer van der Waals (vdW) heterostructures has been demonstrated experimentally only recently[14-19]. Most importantly, it turned out that in practice the atomic-scale Lego works exceptionally well, better than one could have imagined. How it works and why vdW heterostructures deserve attention is discussed in this review.

**Dreamscape**

For any research subject, it is helpful to have a big idea, even if it is unlikely to be realized in its original form. In the case of graphene, its biggest ambition has been to grow into the new silicon, offering a lifeline for Moore's law[1]. At the time of writing, dreams of other 2D crystals are relatively more modest. They are often about offering alternative solutions that compensate for graphene weaknesses[7-11]. In contrast, vdW heterostructures do not lack in ambition.

Imagine the following structure. Graphene is put on top of a dielectric crystal a few layers thick (e.g., mica), and the sequence is repeated again and again. The resulting vdW crystal is superficially similar to superconducting copper oxides where graphene plays a role of conductive CuO planes and the 2D high-κ dielectrics provide interplanar spacing. The critical temperature $T_C$ of oxide superconductors depends on many materials parameters, including CuO interlayer spacing[20,21]. Careful tuning of these parameters allowed $T_C$ above 130K. However, the standard growth techniques offer a limited scope to vary the parameters, and the progress has stalled. What if we mimic layered superconductors by using the atomic-scale Lego? Bismuth strontium calcium copper oxide superconductors (BSCCO) can literally be disassembled into individual atomically thin planes[6]. Their reassembly with some intelligently guessed differences seems worth a try, especially when the mechanism of high-$T_C$ superconductivity remains unknown. Moreover, graphene seems to be a natural 2D component in search for new layered superconductors. Indeed, intercalated graphite exhibits respectable $T_C$ above 10 K (ref. 22) and can be viewed as a stack of heavily doped graphene planes with an increased interlayer distance. What if a dielectric plane of, for example, BSCCO or hBN is added in between the planes of intercalated graphite? Such artificial materials engineered with one atomic plane accuracy would have been a science fiction a few years ago but are within the grasp of today's technology.

We have used the above example for its straightforward appeal and because high-$T_C$ superconductivity in doped graphene has widely been hypothesized (e.g., refs. 23-26). Yet, vdW heterostructures bring to mind not one but many similar speculative ideas. Another example is a room-temperature excitonic superfluidity suggested for two graphene layers separated by an ultra-thin dielectric[27,28]. On the scale of such dramatic perspectives, it may sound fairly modest that vdW heterostructures also offer a helping hand in graphene's efforts to go beyond silicon[1,15]. Of course, one can think of many arguments why these or similar grand ideas can fail. In blue-sky research, even the most plausible scenarios often do. However, 'big dreams' are essential to keep us trying and serve as Ariadne's thread when exploring new topics. Having said that, it is equally important not to get lost on the way and check such ideas against the contemporary reality.

**Layered reality check**

Before explaining how to make vdW heterostructures, it is instructive to review the existing library of 2D crystals, those individual components that can be used in the assembly. In principle, there exist hundreds of layered materials that cleave easily, and one can naturally think of using, e.g., the same Scotch tape technique to isolate their atomic planes[6]. Unfortunately, this is not so straightforward. One must remember that 1) melting temperature ($T$) decreases with decreasing thin films' thickness and 2) most materials survive our ambient conditions only due to natural passivation of their surfaces[29]. Monolayers have two surfaces and no bulk in between, which presents the extreme case of surface science. On the other hand, neither of the procedures developed for isolating 2D crystals can currently be carried out in high vacuum or at low *T*, typical surface science requirements. In this regard, it is informative to mention that graphene monolayers are notably more reactive than even its bilayers[30,31]. In short, many 2D crystals imaginable in theory are unlikely to survive in reality because they would corrode, decompose, segregate, etc.[29]

High thermal and chemical stability of a 3D crystal is essential to even contemplate a possibility of its 2D counterpart. Graphite flaunts both, allowing graphene under ambient conditions. The same is valid for other stable 2D crystals in Table 1. Nonetheless, the surface of $MoS_2$ starts oxidizing in moist air already at 85°C (ref. 32). Even graphene would not survive, if our rooms were twice hotter, at 300°C (ref. 30). Take for example GaSe, $TaS_2$ or $Bi_2Se_3$. These materials are stable in bulk but, being cleaved down to a few layers, they corrode. An illuminating case is silicene[33,34]. This 2D silicon can be grown epitaxially and investigated in high vacuum. However, it won't survive isolation from its parent substrates and in air. With the contemporary technologies, the silicene transistors envisaged in literature cannot be made. These examples are to say that, because of poor stability of atomically thin films compared to their 3D counterparts, the library of 2D crystals should be relatively limited.

Another important consideration is interfacial contamination. Adsorbates such as water, hydrocarbons, etc. cover every surface, unless it is prepared under extreme surface science conditions. Graphene is densely covered with hydrocarbons, even after annealing in high vacuum of a transmission electron microscope. It takes considerable effort to find clean patches of several nm in size[16]. Note that this contamination is highly mobile and usually remains unnoticeable for scanning probe microscopy. If isolated 2D crystals are stacked together, the surface contamination becomes trapped in between layers. Therefore, vdW heterostructures should generally be expected to become 'layer cakes' glued by contamination rather than the neat crystals imagined in Fig. 1. This scenario would take away much of the appeal from vdW heterostructures because 'layer cakes' would result in poor control and reproducibility. Fortunately, it turns out that contamination can clean itself off the interfaces[16,19,35] as further discussed below.

**2D family values**

At the time of writing, we can be certain of the existence of more than a dozen of different 2D crystals under ambient conditions. First of all, these are monolayers of graphite, hBN and $MoS_2$, which have been studied extensively. It is probably not coincidental that these materials are widely used as solid lubricants, which requires high thermal and chemical stability. There are also 2D $WS_2$, $WSe_2$ and $MoSe_2$ which are chemically, structurally and electronically similar to $MoS_2$. Despite little research done so far on the latter monolayers, it is safe to add them to the 2D library, too (Table 1).

Among the above 2D crystals, graphene is an unequivocal champion, exhibiting the highest mechanical strength and crystal and electronic quality. It is likely to be the most common component in future vdW heterostructures and devices. Latest developments on graphene include micrometre-scale ballistic transport at room *T* (refs. 3,36) and low-*T* carrier mobilities $\mu \sim 10^6$ cm$^2$V$^{-1}$s$^{-1}$ in suspended devices[4,5]. The runner-up is 2D hBN, so called "white graphene". Its rise started when bulk hBN crystals[37,38] were shown to be an exceptional substrate for graphene, allowing a 10-fold increase in its electronic quality[2,38]. This advance attracted immediate attention and, shortly after, few- and mono- layers of hBN were used as gate dielectrics[14,39] and tunnel barriers (2D hBN can sustain biases up to ≈0.8V/nm and be free from pin holes)[40,41].

Monolayers of $MoS_2$ were studied earlier[6,42] including the demonstration of the electric field effect but they received little attention until devices with switching on/off ratios of >10$^8$ and room-*T* $\mu \sim$100 cm$^2$V$^{-1}$s$^{-1}$ were reported[12]. Although these $\mu$ are much lower than in graphene, they are still remarkably high compared to thin film semiconductors. The large on/off ratios are due to a sizeable bandgap in $MoS_2$. It is direct in a monolayer (≈1.8eV) whereas bi- and few- layer $MoS_2$ are indirect bandgap semiconductors[43,44]. Semiconductors with a direct gap are of special interest for use in optics and optoelectronics. Further interest in monolayer $MoS_2$ is due to the broken centrosymmetry that allows efficient spin and valley polarization by optical pumping[45,46]. This research is stimulated by availability of large molybdenite crystals from several mining sources. The absence of such supply is probably the reason why 2D $WS_2$, $WSe_2$ and $MoSe_2$ attract relatively little attention despite the fact that Raman and transport studies have revealed their electronic structures and quality similar to that of $MoS_2$ (refs. 47-49). The differences between these dichalcogenides worth noting are stronger spin-orbit coupling in the W compounds and lower stability of the Se compounds.

There have been reports on exfoliation of many other layered chalcogenides down to a monolayer (Table 1). However, we have chosen only to pencil them in the 2D library because little remains known about their stability, let alone optical and transport properties. In some cases, it is even unclear whether the observed flakes and suspensions are indeed 2D counterparts of the parent crystals or present different chemical entities after exposure to air or liquids. In our experience, monolayers of metallic dichalcogenides are unstable in air (T. Georgiou *et al* in preparation).

| graphene family | graphene | hBN 'white graphene' | | BCN | fluorographene | graphene oxide |
|---|---|---|---|---|---|---|
| 2D chalcogenides | $MoS_2$, $WS_2$, $MoSe_2$, $WSe_2$ | | semiconducting dichalcogenides: $MoTe_2$, $WTe_2$, $ZrS_2$, $ZrSe_2$, etc. | metallic dichalcogenides: $NbSe_2$, $NbS_2$, $TaS_2$, $TiS_2$, $NiSe_2$, etc. | | |
| | | | | layered semiconductors: GaSe, GaTe, InSe, $Bi_2Se_3$, etc. | | |
| 2D oxides | micas, BSCCO | $MoO_3$, $WO_3$ | perovskite-type: $LaNb_2O_7$, $(Ca,Sr)_2Nb_3O_{10}$, $Bi_4Ti_3O_{12}$, $Ca_2Ta_2TiO_{10}$, etc. | | hydroxides: $Ni(OH)_2$, $Eu(OH)_2$, etc. | |
| | layered Cu oxides | $TiO_2$, $MnO_2$, $V_2O_5$, $TaO_3$, $RuO_2$, etc. | | | OTHERS | |

Table 1. **Current 2D library.** In blue cells are monolayers proven to be stable under ambient conditions (room *T* in air); green – probably stable in air; pink – unstable in air but maybe stable in inert atmosphere. Grey cells indicate 3D compounds which have been successfully exfoliated down to monolayers as evidenced by, e.g., atomic force microscopy but with little further information. Summarized from refs 6-11,42,50. Note that, after intercalation and exfoliation, the oxides and hydroxides may exhibit stoichiometry different from their 3D parents (e.g., $TiO_2$ exfoliates into a stoichiometric monolayer of $Ti_{0.87}O_2$)[8]. Cell OTHERS indicates that many other 2D crystals including borides, carbides, nitrides, etc. have been[7-11] or can be isolated.

Another group of 2D crystals are numerous oxides including monolayers of $TiO_2$, $MoO_3$, $WO_3$, mica and perovskite-like crystals such as BSCCO and $Sr_2Nb_3O_{10}$ (for review, see refs. 7-11). As oxides, these crystals are less susceptible to air but they tend to lose oxygen and may react with minority chemicals (e.g., water and hydrogen). Similarly to other atomically-thin crystals, properties of 2D oxides are expected to differ from those of their parents due to quantum confinement. Indeed, monolayer oxides often exhibit lower dielectric constants and larger band gaps than their 3D counterparts[8] and can exhibit charge density waves[6]. Unfortunately, information about oxide monolayers is limited mostly to their observation by atomic force and electron microscopies. The heavy artillery of physics and nanotechnology such as the electric field effect, Raman and optical spectroscopy, tunnelling, etc. have not been applied yet to isolated 2D oxides. We should also mention 2D hydroxides that can be exfoliated down to a monolayer but even less is known about them[7].

Finally, there are several graphene derivatives (Table 1). One of them is fluorographene, a stable stoichiometric wide-gap insulator, which can be obtained by fluorination of graphene[51]. Unfortunately, only crystals with poor electronic quality were reported so far[51]. Graphane (fully hydrogenated graphene) gradually loses its hydrogen[31] and is unlikely to be useful for making heterostructures. Nonetheless, note that hydrogenated or other derivatives can sometimes be more stable than 2D crystals themselves[52]. Finally, let us mention graphene oxide[53] and monolayers of boron carbon nitride (BCN)[54,55] which although non-stoichiometric can also be considered for designing vdW heterostructures.

**Rules of survival**

As interest in graphene-like crystals rapidly grows[7-11], search for new 2D candidates is expected to intensify, too. In this regard, the following rule of thumb can be helpful. First, 3D materials with melting *T* >1,000°C have best chances of having 2D counterparts stable at room *T*. Second, 3D parents must be chemically inert and exhibit no decomposed surface layer in air or an alternative environment where exfoliation takes place. Third, insulating and semiconducting 2D crystals are more likely to be stable than metallic ones, owing to generally higher reactivity of metals. In all cases, visual evaluation and Raman spectroscopy are helpful to provide a rapid test for the absence of corrosion and the presence of essential signatures indicating a similarity to the parent crystal. However, the ultimate proof lies with electrical measurements of either in-plane transport for conducting 2D crystals or out-of-plane tunnelling through insulating ones to check for their homogeneity and the absence of pin holes.

As a further step to expand the 2D library, one can perform isolation and encapsulation in an inert atmosphere. Many metallic 2D dichalcogenides may then remain stable at room $T$ as their stability in solvents seems indicate[50,56]. This approach can also lead to higher electronic quality for current favourites such as graphene and 2D $MoS_2$. Exfoliation-encapsulation at low $T$ (e.g., liquid nitrogen) is in principle possible but for the moment too difficult to contemplate for practical use. Lastly, it is important to mention that monolayers may exist without a layered 3D parent (cf. silicene and monolayers of $Y_2O_3$ and ZnO)[57,58]. If the monolayers are sufficiently stable, the substrate can be etched away, as demonstrated for graphene grown on metal foils[59-61]. This can provide access to 2D crystals without 3D layered analogues in nature.

**Lego on atomic scale**

It is no longer adventurous to imagine an automated, roll-to-roll assembly[61] of vdW heterostructures by using sheets of epitaxially grown 2D crystals[59-63]. However, concerted efforts towards such assembly are expected only when a particular heterostructure proves to be worth of attention, as happened for the case of graphene on hBN[64]. For scouting the research area, manual assembly is likely to remain the favourite approach. It offers high throughput and relatively easy changes in layer sequences. Likewise, individual 2D compounds are to continue being obtained by the Scotch tape technique that has so far provided crystallites of unmatched quality. Nevertheless, we expect an increasingly frequent use of epitaxially grown graphene, 2D hBN, 2D $MoS_2$, etc. for making proof-of-concept vdW heterostructures.

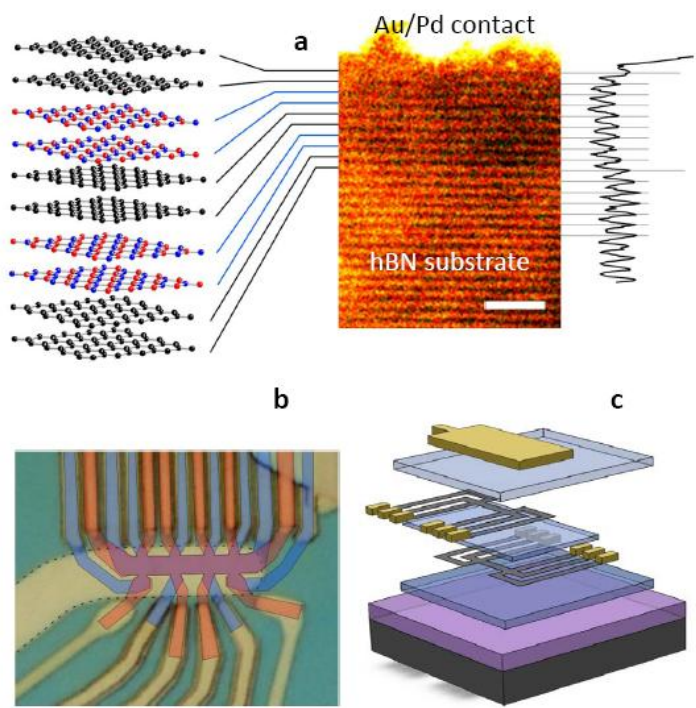

Fig. 2. **State of the art vdW structures and devices**. **a** – Graphene-hBN superlattice consisting of 6 stacked bilayers. Right: Its cross-section and intensity profile as seen by scanning transmission electron microscopy. Left: schematic view of the layer sequence. The topmost hBN bilayer is not visible, being merged with the metallic contact. Scale bar, 2nm. Adapted from ref. 16. **b,c** – Double-layer graphene heterostructures[18]. An optical image of a working device (b) and its schematics (c). Two graphene Hall bars are accurately aligned, separated by a trilayer hBN crystal and encapsulated between relatively thick hBN crystals. Colours in (b) indicate the top (blue) and bottom (orange) Hall bars and their overlapping region in violet. The graphene areas are invisible in the final device image because of the top Au gate outlined by dashes. The scale is given by the width of the Hall bars, 1.5 μm.

At the time of writing, only a few groups reported vdW heterostructures made from more than two atomically thin crystals, and only graphene and few-layer hBN, $MoS_2$ and $WS_2$ were used for this assembly[14-19,65,66]. A typical stacking procedure starts with isolating micrometre-sized 2D crystals on top of a thin transparent film (e.g., polymer). The resulting 2D crystal provides one brick for the Lego wall in Fig. 1 and can now be put face down onto a chosen target. The supporting film is then removed or dissolved. More 2D crystals are produced, and the transfer is repeated again and again, until a desired stack is assembled. Conceptually, this is simple and requires only basic facilities such as a good optical microscope. In practice, the fabrication takes months to master. In addition to the standard cleanroom procedures (cleaning, dissolving, resist spinning, etc.), positioning of different 2D crystals with micrometre accuracy over each other is necessary. This is done under the microscope by using micromanipulators. The crystals must be put in soft contact without rubbing and, preferably, no liquid or polymer is allowed in contact with cleaved surfaces to minimize contamination. Thermal annealing in an inert atmosphere can often be helpful after adding each new layer. For transport measurements, 2D crystals are plasma etched into, e.g., Hall bars with contacts evaporated as the final step.

Despite dozens of involving steps, sophisticated multilayer structures can now be produced within a matter of days. Figure 2 shows two such examples. One is a vdW superlattice made from six alternating bilayers of graphene and hBN. This is the largest number of 2D crystals in a vdW heterostructure, which has been reported so far[16]. On the other hand, the most challenging design has probably been double-layer graphene devices[18] such as shown in Fig. 2b-c. Let us emphasize that interfaces in these heterostructures are found to be clean and atomically sharp[16,19], without the contaminating 'goo' that always covers 2D crystals even in high vacuum (see the reality check section). The reason for the clean interfaces is the vdW forces that attract adjacent crystals and effectively squeeze out trapped contaminants or force them into micrometre-sized 'bubbles'[16]. This allows 10-μm-scale devices practically free from contamination. It is interesting to note that atomically sharp interfaces are practically impossible to achieve by other techniques, including molecular beam epitaxy, because of the island growth.

**Little evolutionary steps**

Despite the recognized availability of various isolated 2D crystals[6], practical steps towards their vdW assembly were taken only after 2010. An important stimulus was the demonstration that hBN could serve as a high-quality substrate for graphene[2] (many other substrates including pyrolytic hBN were tried before but unsuccessfully[35,67]). This led to rapid development of transfer procedures. The next logical step was encapsulation where thin hBN crystals served not only as the substrate but also as a protective cover for graphene[3]. The encapsulation has proven its virtue by enabling devices with consistently high quality that does not deteriorate under ambient conditions. The currently achieved $\mu$ for graphene on a substrate are ~100,000 $cm^2V^{-1}s^{-1}$ but can reach up to 500,000 $cm^2V^{-1}s^{-1}$ at low $T$. Such high quality can be witnessed directly as negative bend resistance and magnetic focusing[3,36] (Fig. 3a). These ballistic effects persist up to room $T$. The encapsulation also results in high spatial uniformity so that capacitors of >100 $\mu m^2$ in size exhibit quantum oscillations in magnetic fields $B$ as low as $\approx$0.2T (ref. 68).

The next evolutionary step has been 'vertical' devices in which few-layer thick crystals of hBN, $MoS_2$ or $WS_2$ were used as tunnel barriers with graphene serving as one or both electrodes[15,19]. These devices required 3 to 4 transfers but no plasma etching to define geometry. Although sensitive to charge inhomogeneity, vertical devices usually do not pose critical demands on $\mu$. The tunnelling heterostructures allowed the demonstration of a new kind of electronic devices, field-effect tunnelling transistors[15]. In this case, tunnel current is controlled by changes in the electrode's Fermi energy, which could be varied by gate voltage by as much as $\approx$0.5 eV due to the low density of states in monolayer graphene. An increase in the Fermi energy effectively lowers the tunnel barrier, even if no bias is applied[15]. This is in contrast to the standard Fowler-Nordheim mechanism that is based on tilting of the top of the tunnel barrier by applied bias. The vdW tunnelling devices exhibit on-off switching ratio >$10^6$ at room $T$ (refs. 15,19,69).

A higher level of complexity is presented by the graphene-hBN superlattice shown in Fig. 2a. It proves the concept that thin films of new 3D materials consisting of dozens of atomic layers are in principle possible by reassembly, as discussed in the Dreamscape section. In the case of Fig. 2a, hBN bilayers served as spacers whereas bilayer graphene (rather than its monolayer) was chosen to facilitate intercalation to reach a high density of states. Further efforts in making and investigating such multilayer structures are expected due to the interest generated by a large amount of literature on possible collective phenomena in graphene-based systems[23-28,70-72].

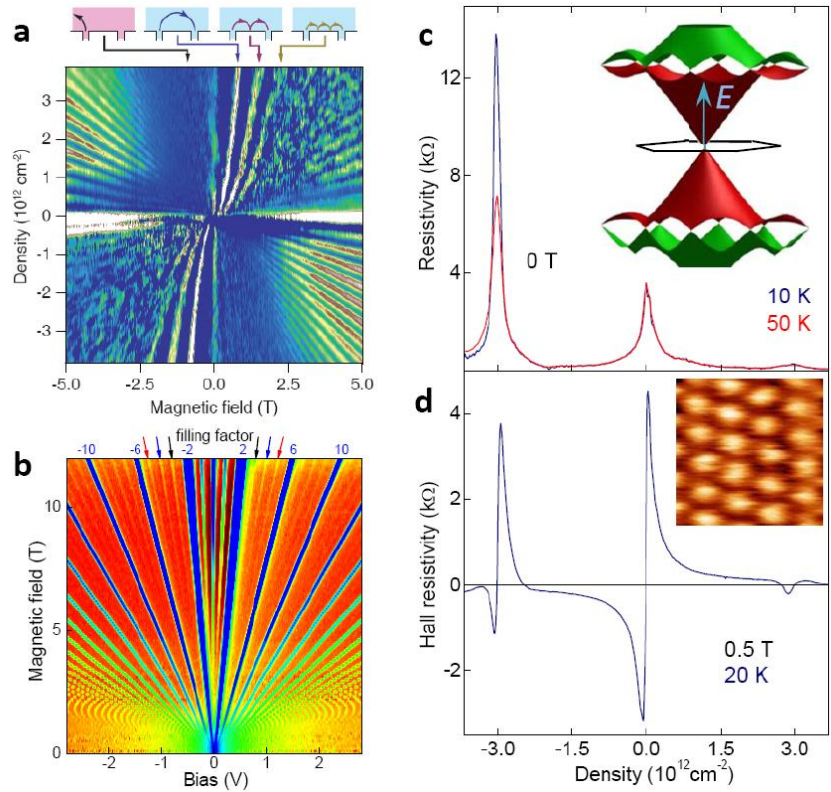

Fig. 3. **Early harvest in vdW fields**. **a** – Magnetic focusing in graphene on hBN. Pronounced resonances are observed if the size of a cyclotron orbit becomes commensurate with the distance between narrow graphene leads used as an injector and detector (adapted from ref. 36). Bright colours show maxima in conductivity as a function of carrier density and *B*. The upper panel illustrates the corresponding trajectories. **b** – Quantum capacitance of encapsulated graphene as a function of gate voltage and *B*. In this device, spin and valley degeneracies are lifted above 10T. Adapted from ref. 68. **c,d** – Importance of crystallographic alignment. The standard Dirac-like spectrum is strongly reconstructed for graphene on hBN, and new Dirac cones appear in both valence and conduction bands [inset in (c)]. This leads to pronounced peaks in resistivity (c) and the Hall effect changes sign (d). Inset in (d) shows the moiré patterns that lead to the spectral reconstruction. Adapted from ref. 76.

The double-layer devices in Fig. 2 are the latest state of the art for vdW heterostructures. They were designed to probe in-plane transport in the regime of ultimately strong electron-electron interaction between electrically isolated 2D systems[18,73]. The separation of the graphene layers can be as small as 3 hBN layers (≈1 nm)[18] but this still provides a sufficiently high potential barrier to suppress electron tunnelling. The layers continue to 'feel' each other strongly through Coulomb interactions. The 1nm separation is much smaller than the in-plane distance between charge carriers in graphene, which is typically ~10nm and nominally diverges near the neutrality point. This makes the interlayer separation the smallest spatial parameter in the problem. Therefore, the two electronic liquids in double-layer graphene effectively nest within the same plane, but still they can be tuned and measured separately.

The ambition behind the double-layer heterostructures is to address many possible collective states including Wigner crystallization, excitonic superfluidity, itinerant magnetism, etc. Such many-body phenomena have so far been the domain of low-*T* physics. In particular, Bose condensation of interlayer excitons has been reported at sub-kelvin *T* in double-layer semiconductor heterostructures, and Coulomb drag was used as a way to detect the condensate[74,75]. The hope was that double-layer graphene would allow a similar superfluid state but at much higher temperature[27,28] because in this case the

Coulomb energy of interlayer interaction can be as high as >0.1eV. The measurements[18] revealed many interesting and some unexpected features but no sign of superfluidity in zero $B$ so far. By analogy with the semiconductor heterostructures[74,75], the regime of quantizing $B$ remains most promising for the observation of coherent electronic states, but this was not investigated yet.

A further principal step in sophistication of vdW heterostructures was to include crystallographic alignment, which can be done with accuracy of <1° (ref. 76). Although the interaction between stacked 2D crystals is relatively weak (~10meVÅ$^{-2}$; ref. 77), electron orbitals still extend out of the plane and affect charge carriers in an adjacent 2D crystal. This influence results in moiré patterns that depend on the rotation angle between joining crystals and their lattice mismatch[78-80] (Fig. 3d). In the case of graphene on hBN, a periodic potential created by the hBN substrate results in additional Dirac cones at high electron and hole densities[76,81] (Fig. 3c). The superlattice effects are remarkably strong so that the Hall effect changes its sign (Fig. 3d) and the new Dirac cones exhibit their own sets of Landau levels[76,81]. This is the best proof so far that interfacial contamination can be negligible and, also, shows that electronic spectra of vdW heterostructures can be tuned by using moiré potentials. As another example of such band gap engineering, one can consider the alignment of monolayers that have close lattice constants (e.g., $MoS_2$ and $WSe_2$). The resulting vdW crystal is expected to exhibit optical and electronic properties distinct from its components[82].

Many other types of vdW structures and devices are expected to be demonstrated soon, initially by using only a small number of 2D crystals. Among obvious objectives are various proximity effects. To this end, 2D crystals can be put on top of atomically-flat crystals exhibiting magnetism, ferroelectricity, spin-orbit coupling, etc. For example, graphene encapsulated in $WS_2$ is likely to exhibit an induced spin-orbit interaction that should affect transport properties and may push the system into the topological insulator state.

**Handicraft on industrial scale**

The growing interest in vdW heterostructures is not limited to new physics and materials science. There is a massive potential for applications, too. Here we avoid early speculations because the applied interest in vdW heterostructures can already be justified if one considers them as a way to accelerate those myriads of applications offered by graphene itself[1]. The recently demonstrated new graphene-based device architectures[1,15,83] provide straightforward examples from this perspective.

Any industrial application will obviously require a scalable approach to the vdW assembly. To this end, significant efforts have been reported to epitaxially grow graphene, 2D hBN and 2D $MoS_2$ on top of each other[10,84-88]. However, it is a daunting task to find the right conditions for so-called vdW epitaxy[89] because the weak interlayer interaction generally favours the island growth rather than continuous monolayers. Another scalable approach is layer by layer deposition from 2D-crystal suspensions by using Langmuir-Blodgett or similar techniques[8,90]. One can also mix suspensions of different 2D crystals and then make layer-by-layer laminates, relying on self-organizational assembly (flocculation)[11]. Unfortunately, micrometre-sized crystallites in suspensions cannot provide large continuous layers, and this would limit possible applications for such vdW laminates. Currently, they are considered for use as designer ultra-thin dielectrics[8], selectively-permeable membranes[91] and composites materials[92].

At the time of writing, the most feasible approach to industrial scale production of vdW heterostructures seems to grow individual mono- and few- layers on catalytic substrates, then isolate and transfer these 2D sheets on top of each other. This route has already been proven to be scalable[61,64]. If a particular heterostructure attracts sufficient interest in terms of applications, it seems inevitable that its production can be scaled up by trying a variety of available approaches.

**Long live graphene**

After many years of intensive effort, graphene research should logically have reached a mature stage. However, the possibility to combine graphene with other 2D crystals has expanded this field dramatically, well beyond simple graphene or 2D $MoS_2$. The interest in vdW heterostructures is rising as quickly as it happened to graphene a few years ago. As technology of making vdW heterostructures moves from its humble beginnings, increasingly sophisticated devices and materials should become available to more and more research groups. This is likely to cause a snowball effect because, with so many 2D crystals, sequences and parameters to consider, the choice of possible vdW structures is limited only by our imagination. Even with the 2D components that had been proven stable, it will take time and effort to explore the huge parameter space. Decades-long research on semiconductor heterostructures and devices may serve as a guide to judge on probable longevity of research on vdW materials, beyond simple graphene.

**Acknowledgements** We thank all participants of the Friday Graphene Seminar in Manchester for discussions. This work was supported by the Royal Society, the European Research Council, the Körber Foundation, the Office of Naval Research and the Air Force Office of Scientific Research.